\documentclass[conference]{IEEEtran}
\ifCLASSINFOpdf
\else
\fi
\usepackage[colorinlistoftodos]{todonotes}
\usepackage{amsmath}

\usepackage{placeins}
\usepackage{dirtree}

\usepackage{floatrow}
\usepackage{blindtext}

\begin{document}
%
\title{SLA Violation Prediction In Cloud Computing:\\
A Machine Learning Perspective}

\author{\IEEEauthorblockN{Reyhane Askari Hemmat}
\IEEEauthorblockA{Departement d'informatique et\\
de recherche op\'{e}rationnelle\\
Universit\'{e} de Montr\'{e}al  \\
Email: reyhane.askari.hemmat@umontreal.ca}
\and
\IEEEauthorblockN{Abdelhakim Hafid}
\IEEEauthorblockA{Departement d'informatique et\\
de recherche op\'{e}rationnelle\\
Universit\'{e} de Montr\'{e}al  \\
Email: ahafid@iro.umontreal.ca}}


%


\maketitle

\begin{abstract}
Service level agreement (SLA) is an essential part of cloud systems to ensure maximum availability of services for customers. With a violation of SLA, the provider has to pay penalties. Thus, being able to predict SLA violations favors both the customers and the providers. In this paper, we explore two machine learning models: Naive Bayes and Random Forest Classifiers to predict SLA violations. Since SLA violations are a rare event in the real world ($ \sim 0.2\%$), the classification task becomes more challenging. In order to overcome these challenges, we use several re-sampling methods such as Random Over and Under Sampling, SMOTH, NearMiss (1,2,3), One-sided Selection, Neighborhood Cleaning Rule, etc. to re-balance the dataset. We use the Google Cloud Cluster trace as the dataset to examine these different methods. We find that random forests with SMOTE-ENN re-sampling have the best performance among other methods with the accuracy of 0.9988\% and $F_1$ score of 0.9980. 

\end{abstract}


%
\IEEEpeerreviewmaketitle

\section{Introduction}
The usage of cloud systems have been so prevalent that it is hard to picture to use many services and applications without cloud computing. Cloud computing reduces the maintenance costs of the service and also allows users to access on demand services without being involved in technical implementation details. The relationship between a cloud provider and a customer is governed with a \textit{Service Level Agreement} (SLA) that is established to define the level of the service and its associated costs. SLA usually contains specific parameters and a minimum level of quality for each element of the service that is negotiated between the provider and the customer \cite{1}.

An SLA is an important part of each contract because a provider would like to allocate the least amount of resources for each customer to reduce the cost of its server infrastructure. At the same time, the provider needs to avoid having penalties due to failure of providing the agreed service. The failure of providing a service is called an \textit{SLA violation}. On the other hand, the customer would like to receive the service on demand and without any interruptions. Despite these high availability rates, violations do happen in real world and have caused both the provider and the customer heavy costs.

According to \cite{2}, SLA management has six phases: SLA contract definition, basic schema with the \textit{Quality of Service} (QoS) parameters, SLA negotiation, SLA monitoring, SLA violation detection and SLA enforcement. An essential part of SLA monitoring is to be able to predict violations enabling providers to reallocate the resources accordingly before occurrence of violations. Moreover, from a customer's point of view, a trusted provider can be chosen based on the provider's future violations.

In this paper, through a set of experiments, we identify best performing Machine Learning models which are able to predict SLA violations on a real world dataset. It is worth mentioning that due to skewness of real world datasets\footnote{For example, Google Cloud Cluster has 99.90$\%$ availability versus 0.10$\%$ violations.}, it is a challenging problem to predict violations. We report our results on a subset of the Google Cloud Cluster trace dataset \cite{3}. The results presented in Section \ref{sec:results} show that the best performance is achieved using the Random Forest \cite{4} method with an accuracy of 99.88$\%$.

The remainder of this paper is organized as follows. Section II describes the previous works on SLA violation prediction. Section II describes our dataset which was based on Google Cluster Trace published on 2011. Section IV defines SLA violation and how we can find them in the dataset. Section V explains the two models that we used for the prediction task. Section VI describes our experimental setup including how we overcome data skewness and what kind of error metrics are used. In section VII we discuss the our results and finally, Section VIII concludes the paper and presents future work.

\section{Related Works}
Many models have been proposed in recent years to tackle SLA management problems. Imran et al. \cite{5} use a map-reduce model to detect violations and find the most probable causes of SLA violations using Holt-Winters forecasting. In \cite{6}, the authors proposed an SLA aware resource scheduling model that determines how to allocate coming requests without explicit prediction of violations.
Similarly, in \cite{7} , Mohammed et al. proposed an SLA-based trust model for cloud that selects the provider based on a selection scheme; in this scheme, although no violation is predicted, the customers are grouped according to business needs and the most trusted cloud provider is selected based on the customer's non-functional requirements.

In a similar work for predicting SLA violations in composite services, in \cite{8} the authors propose a regression machine learning model; the regression model is implemented using the WEKA framework which cannot be scaled to real world environments where the scale of the dataset is much bigger compared to the one which is used in our paper.

In \cite{9}, the authors proposed a model for predicting \textit{host load} using real data of Google Compute Cluster. A Bayesian model is used to predict the mean load over long-term and consecutive future time intervals. Although the load could be predicted, there is no detection of SLA violations in the paper.

The authors in \cite{10} propose a provisioning method that monitors and also predicts 
future loads. Based on the predicted loads, an autonomous elasticity controls the number of allocated virtual machine to a job. While no explicit detection of violations are predicted, the number of delayed requests was lowered by a factor of three. 

The authors in \cite{11}, use unsupervised learning to cluster the resource usage and duration of services to avoid violations of Google Cluster trace dataset. If a violation happens inside a cluster of services, the other services inside the cluster will be assigned resources to avoid the violation. This helps in violation avoidance in the cluster but there is no specific prediction of SLA violations for each service. 

The authors in \cite{12} use a Naive Bayes model to predict SLA violations. Despite its good performance, the dataset is generated using simulation. It contains 40\% violations and neglects the fact that in real world, violations are very rare (0.1\%).

\section{Dataset}
\begin{figure}[!ht]
\label{fig:erd}
  \centering
      \includegraphics[width=1.1\textwidth]{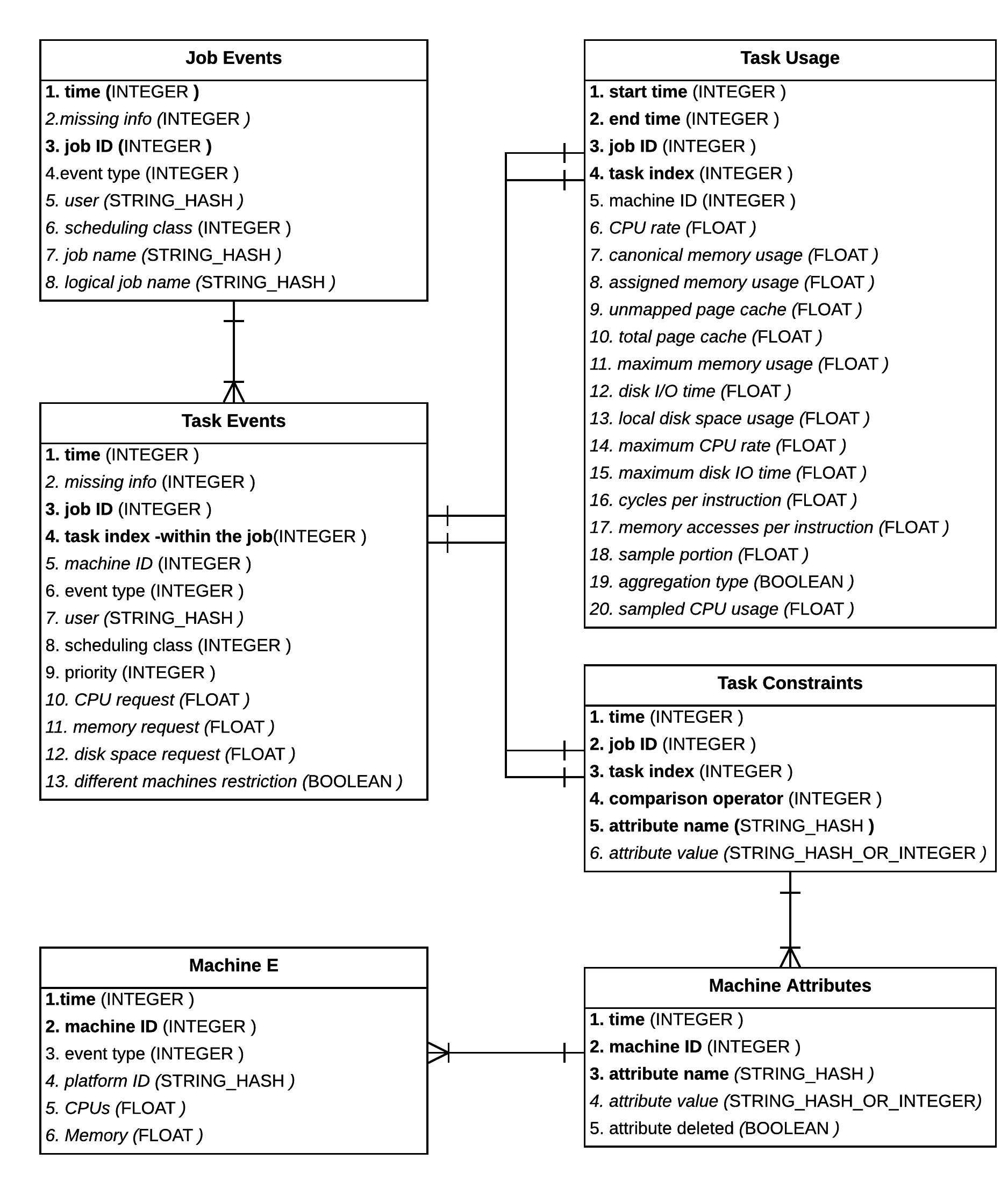}
  \caption{Google's cluster trace dataset ERD. The dataset contains the above five different tables. This ERD is used to define and find violated tasks based on the definition in section \ref{sec:violation}}
\end{figure}

The dataset that we report our results on contains 29-day trace of Google's Cloud Compute. For security reasons, part of the trace has been omitted or obfuscated. For example, the values for CPU, disk and memory have been rescaled by dividing each value by their corresponding largest value in the trace. Also the names of the users' applications have been hashed. The trace has six separate tables: \textit{Job Events}, \textit{Task Events}, \textit{Task Usage}, \textit{Machine Events}, \textit{Machine Attributes}, and \textit{Task Constraints}. The entity relationship diagram of the database is shown in Figure \ref{fig:erd}. 

\begin{figure}[!ht]
  \centering
      \includegraphics[width=0.75\textwidth]{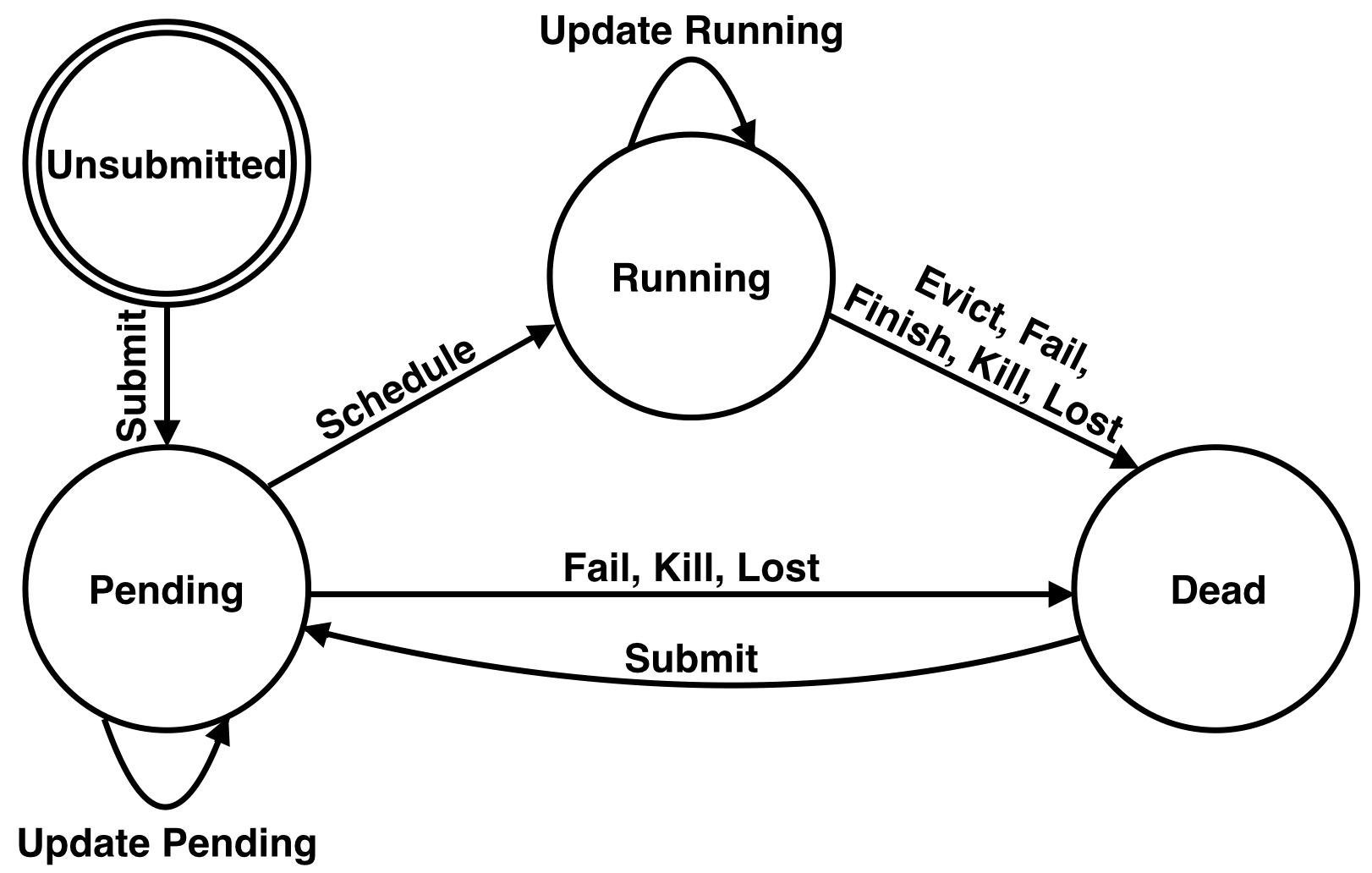}
  \caption{The state transition diagram of a task on Google Cluster machines\cite{3}.}
\label{fig:state_transition}
\end{figure}

User's application submits its required resources as jobs to the cluster and each job is consists of several tasks. The state transition diagram of jobs and tasks is depicted in Figure \ref{fig:state_transition}. The \textit{Job Events} table traces the event cycle of the jobs that were submitted to the cluster. The tasks inside each job are tracked in \textit{Tasks Events} table. Each task is then assigned to a specific machine. \textit{Machine Events} table shows removal or addition of a machine to the cluster or update of its resources. \textit{Machine attributes} table shows the attributes of each machine such as kernel version, clock speed and presence of an external IP address\cite{3}. Tasks can have constraints on machine attributes which are recorded in the \textit{Tasks Constraints} table. 

Information such as \textit{requested CPU}, \textit{requested memory}, \textit{requested disk space}, \textit{scheduling class} and \textit{priority} of the task are all recorded inside tasks events table. The \textit{Task Usage} table contains the actual usage of resources for each task. It contains information such as \textit{assigned memory} and \textit{memory usage}. 

\begin{figure}[!ht]
  \centering
      \includegraphics[width=0.8\textwidth]{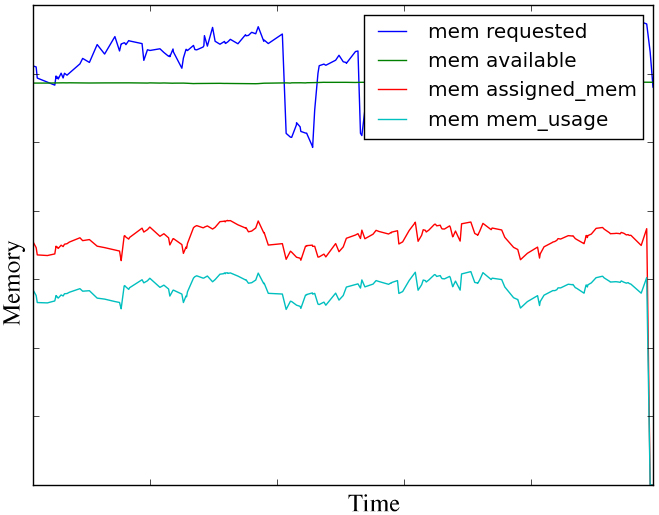}
  \caption{500 snapshots of the requested, available, assigned and used memory of the cluster. The available memory has not changed throughout time and it is usually lesser than the requested memory. On the other hand, the actual usage of the memory is less than available and assigned memory. Google scheduler has reserved a safe margin between the assigned memory and usage of memory.}
\label{fig:snap}
\end{figure}

Figure \ref{fig:snap} illustrates the mechanism of resource allocation in Google's Cluster. It shows the state of the cluster at 500 random \textit{snapshots}. We define a \textit{snapshot} as a moment in time when the total requested resources is calculated. Similarly, available or allocated resources are calculated at each snapshot. In Figure  \ref{fig:snap} the total requested memory, assigned memory, memory usage and available memory of the cluster at each snapshot is calculated using the Task Events, Task Usage and Machine Events tables. It is the nature of cloud to allocate less resources than requested resources and even accept more requests than its available resources. Figure \ref{fig:snap} shows that at all of the 500 snapshots of the cluster, the requested memory to the cluster is much higher than the actual usage of memory.

Another important characteristic of our data set is its skewness. The task of violation detection can be simply considered as a classification problem where we want to predict whether at a specific time in the future the provider will have violation or not. Since the availability rate is very high (97.8\%) and violations do not happen most of the time, the engine have the tendency of always predicting the violation as false. There has been some methods in machine learning to handle such data such as over sampling and under sampling; also it has been recommenced to use generative models\cite{13}.

\section{SLA Violation Definition}
\label{sec:violation}
In order to identify SLA violations we need to have specific details of QoS parameters and Service Level Objectives (SLOs). SLOs are quantitative parameters of an SLA such as availability, throughput and response time. Although we do not have access to the details of SLA for this dataset, we can find violations in the availability of the service using the trace. 

Figure \ref{fig:state_transition} shows the state transition diagram for jobs and tasks in the trace. We define a violation in the availability of the service when a task is evicted and never re-scheduled after that. According to the documentation of the trace \cite{3}, eviction of a task is due to "overcommiting of the scheduler or because the machine on which it was running became unusable (e.g. taken offline for repairs), or because a disk holding the task’s data was lost." Thus, all the tasks that were evicted and never re-scheduled were detected as cases of violations. The percentage of not evicted tasks to the total tasks submitted to the cluster is 97.8$\%$. Thus, the cluster has only 2.2\% violations. Our goal is to use this data and predict future violations. Features such as the amount of requested CPU, disk and memory of the violated tasks and also the available resources at the time of request can be studied to predict future violations.

\section{Prediction Models}
We formulate SLA violation detection as follows: Given a set of features extracted from traces of the cluster, what is the probability of failure\footnote{failure is defined in the section \ref{sec:violation}.} of a submitted task.

To work towards this goal, we establish a classification model and use two algorithms as the core classifier: \textit{Naive Bayes models} and \textit{Random Forest Model}.

\subsection{Naive Bayes Models}
From a probabilistic point of view, the conditional probability of class $k$ among $K$ different classes given a vector representation of $n$ distinct features $\textbf{x} = \{x_1, ..., x_n\}$ can be written as $P(C_k|\textbf{x})$. According to the Bayes theorem \cite{bayes}, the above probability can be reformulated as follows:
\begin{equation}
P(C_k|\textbf{x}) = \frac{P(\textbf{x}|C_k) P(C_k)}{P(\textbf{x})},
\end{equation}
in which $P(C_k|\textbf{x})$ is called the \textit{posterior} meaning our updated knowledge conditioned on the observed data. Two probabilities $P(\textbf{x}|C_k)$ and $P(C_k)$ are called the \textit{likelihood} and the \textit{prior} respectively.

In a classification setup, the denominator $P(\textbf{x})$ is constant. In practice, training such a Bayesian classifier amounts to maximizing the nominator for the target class and minimizing it for the other classes. The nominator is the joint probability of features and classes $P(C_k, x_1, ..., x_n)$ which according to the chain rule, can be reformulated as follows:
\begin{align*}
P(C_k, x_1, ..., x_n) = &P(C_k)*\\
&P(x_1|C_k)*\\
&P(x_2|x_1, C_k)*\\
&P(x_3|x_2, x_1, C_k)*\\
&...\\
&P(x_n|x_{n-1}, ..., x_1, C_k).
\end{align*}

Consequently, since $P(C_k | x_1, ..., x_n) \propto P(C_k, x_1, ..., x_n)$, a classifier can be defined as follows:
\begin{align*}
\hat{c} = \underset{k\in\{1,...,K\}} {\mathrm{argmax}} P(C_k, x_1, ..., x_n).
\end{align*}

\begin{figure}[!ht]
  \centering
      \includegraphics[width=0.5\textwidth]{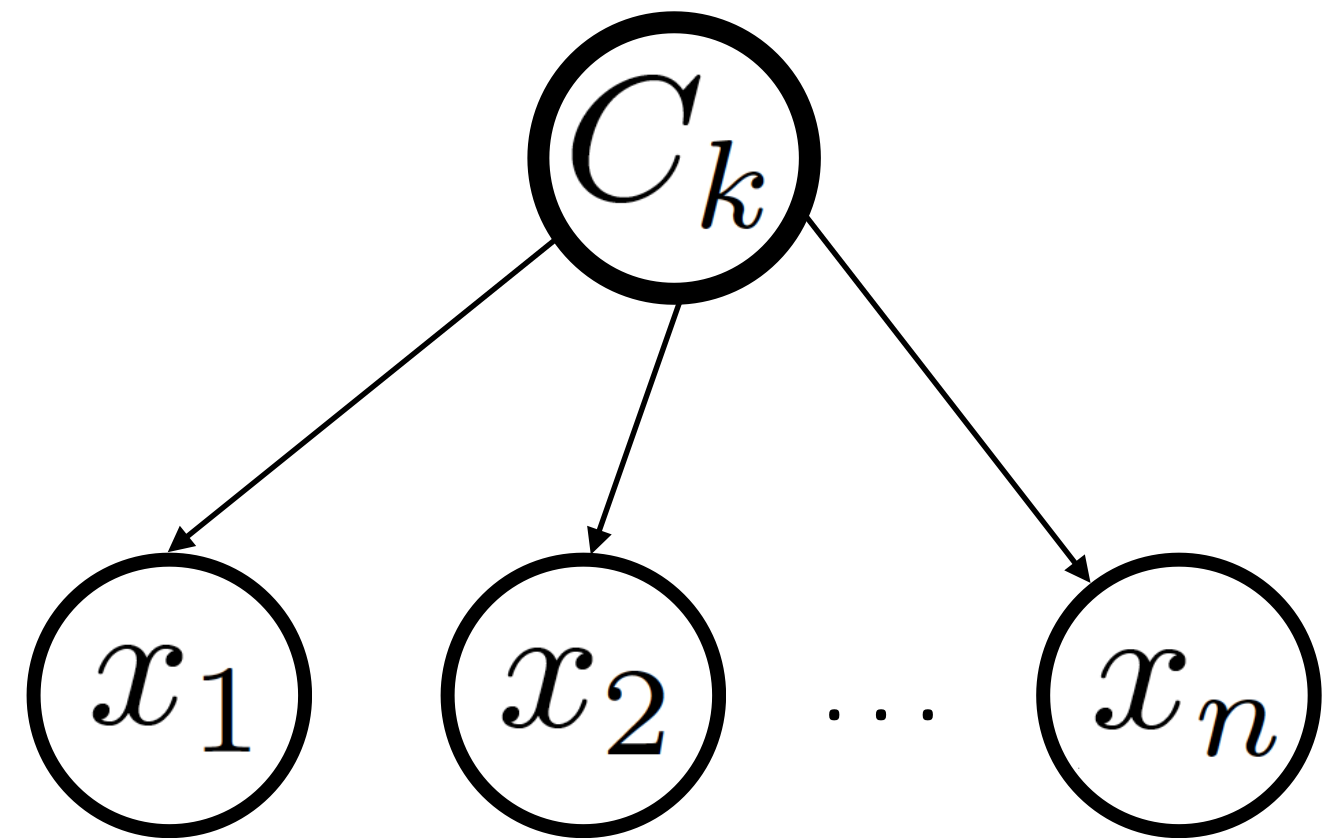}
  \caption{Bayesian network representation of the naive Bayes classifier. According the the graph representation, conditioned on the class $C_k$, $x_i$'s are independent of each other.}
\label{fig:nb}
\end{figure}

In practice, for large number of features, $n$, it is challenging to train such a classifier. One of the simple, yet effective probabilistic classifiers is known as \textit{Naive Bayes}. \textit{Naive Bayes} algorithm has an assumption that given the class label $C_k$, all the features $\{x_1, ..., x_n\}$ are independent of each other. The adjective \textit{naive} comes from the fact that the assumption of class conditional independence is simplistic. A graphical illustration of the this classifier is shown in Figure \ref{fig:nb}.

\subsection{Random Forest Model}
\textit{Decision Tree} is the main building block of the Random Forest model. As a result, we first briefly introduce Decision Tree and then we explain the Random Forest.
\subsubsection{Decision Tree}
\textit{Decision Tree} is a family of scalable classifiers that enjoys the advantage of human-interpretable results. Formally, a classification decision tree is a tree in which each leaf represents a target class, each internal node represents a condition, and each branch corresponds to the outcome of the condition in the parent node.

\begin{figure}[!ht]
  \centering
      \includegraphics[width=0.9\textwidth]{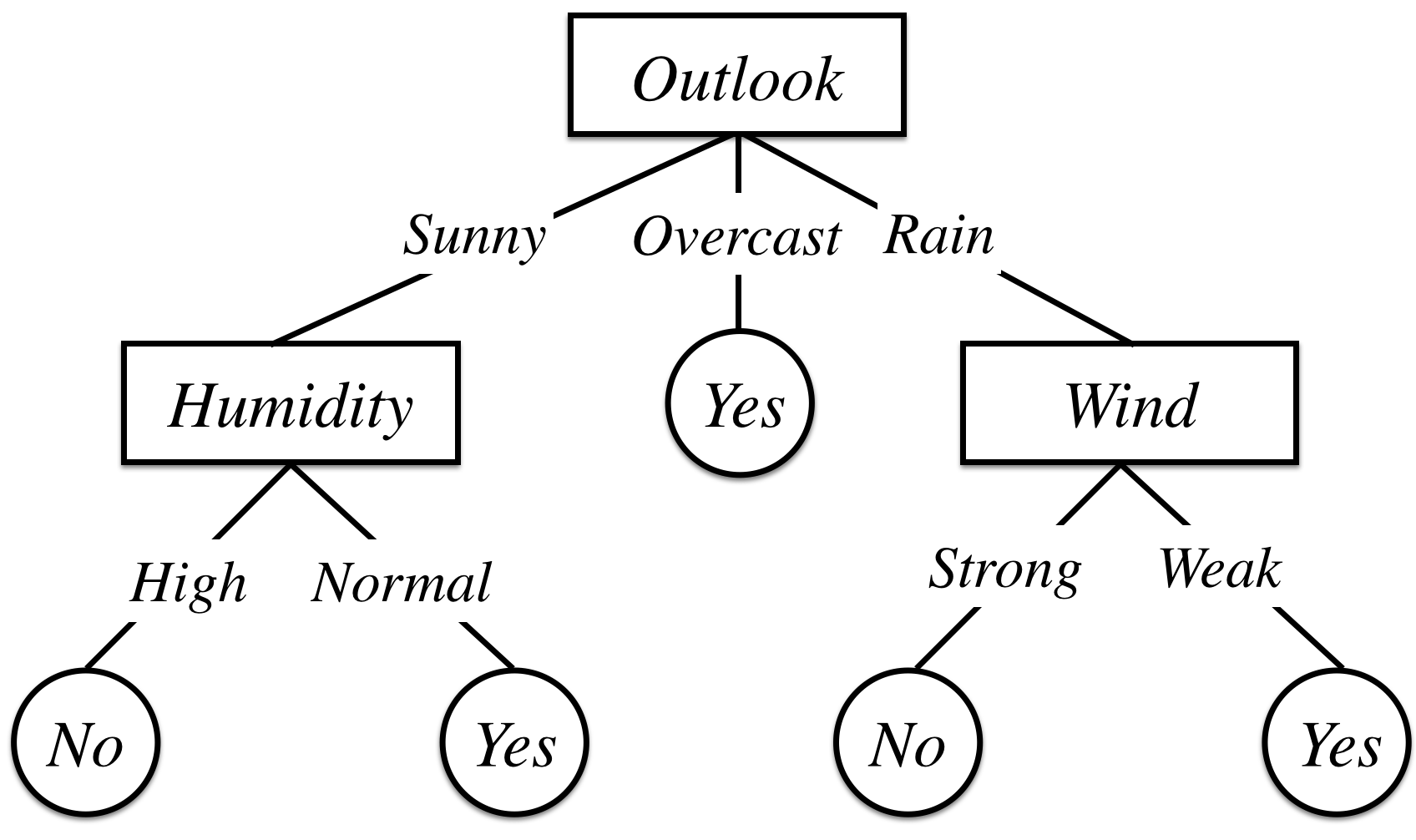}
  \caption{A graphical illustration of a Decision Tree: Classification starts from the top node towards leaves by testing the \textit{Outlook}. After moving to one of the left or the right subtrees, a test on \textit{Humidity} or \textit{Wind} determines the class label. }
\label{fig:dt}
\end{figure}

As a simple example, consider a set of features $\{Outlook, Humidity, Wind\}$ and the target is \textit{PlayTennis} which takes values of \textit{``Yes''} or \textit{``No''}. A trained decision tree is shown in Figure \ref{fig:dt}.

\subsubsection{Decision Tree Learning}
Construction of a decision tree amounts to finding the appropriate conditions as nodes and ordering them from root to the leaves. Conditions are a test on one of the features of the given datapoint. Among different types of tests, we use \textit{Gini Impurity} on each feature as the criterion that splits nodes to their children. \textit{Gini impurity} measures the probability of being wrongly classified for a random datapoint, if the classification is based on the distribution of the targets.

\subsubsection{Random Forrest}
From a geometrical point of view, a decision tree leads to a hierarchical portioning over the feature space. Starting from the top most node in the tree, each node divides the feature space into two or more partitions. Consequently, as the tree gets deeper, more complicated partitioning is done. However, in the case of over-fitting, the partitioned space is over complicated that yields to small error on the training data while a relatively larger error on the test data.

One of the successful ways to overcome the issue of over-fitting is \textit{Random Forrest}. \textit{Random Forrest} amounts to using an ``ensemble'' of decision trees and aggregating their results in order to get a more robust prediction.

Each of the decision trees in random forest is constructed on a subset of the data that is achieved by sampling with replacement from the original dataset. For each decision tree, a different random subset of features is used. In the final stage and aggregation bagging used. 

\section{Experimental Setup}
In our experiments, we fed the historical data to Naive Bayes and Random Forest machine learning models to predict future violations. The task of prediction is modeled as a classification task where we have two classes. Class zero (\textit{violation=0}) is the case of unviolated tasks and class one (\textit{violation=1}) is the case of violated tasks. Since the availability rate is very high (97.8\%), our classes are highly unbalanced. This is known as skewness in dataset which makes the classification task very hard because the classifier will always have the tendency to predict the dominant class.

\subsection{Overcoming Data Skewness}
As depicted in Figure \ref{fig:schematic}, we have experimented several models to re-sample the database and create a more balanced dataset: \textit{Random over-sampling, Random Under-sampling, SMOTE (Synthetic Minority Over-sampling Technique) \cite{14}, Near Miss 1, 2 and 3, One-sided Selection, and Neighborhood Cleaning Rule.}

\thisfloatsetup{%
objectset=raggedright,
}
\begin{figure}[t]
\dirtree{%
.1 Methods.
.2 Random Under Sampling.
.3 Naive Bayes.
.4 Gaussian \dotfill\begin{minipage}[t]{1.0cm}
0.656{.}\end{minipage}.
.4 Bernoulli \dotfill\begin{minipage}[t]{1.0cm}
0.594{.}\end{minipage}.
.3 Random Forest \dotfill\begin{minipage}[t]{1.0cm}
\textbf{0.958}{.}\end{minipage}.
.2 Near miss 1.
.3 Naive Bayes.
.4 Gaussian \dotfill\begin{minipage}[t]{1.0cm}
0.743{.}\end{minipage}.
.4 Bernoulli \dotfill\begin{minipage}[t]{1.0cm}
0.663{.}\end{minipage}.
.3 Random Forest \dotfill\begin{minipage}[t]{1.0cm}
\textbf{0.817}{.}\end{minipage}.
.2 Near miss 2.
.3 Naive Bayes.
.4 Gaussian \dotfill\begin{minipage}[t]{1.0cm}
0.736{.}\end{minipage}.
.4 Bernoulli \dotfill\begin{minipage}[t]{1.0cm}
0.656{.}\end{minipage}.
.3 Random Forest \dotfill\begin{minipage}[t]{1.0cm}
\textbf{0.815}{.}\end{minipage}.
.2 Near miss 3.
.3 Naive Bayes.
.4 Gaussian \dotfill\begin{minipage}[t]{1.0cm}
0.926{.}\end{minipage}.
.4 Bernoulli \dotfill\begin{minipage}[t]{1.0cm}
0.954{.}\end{minipage}.
.3 Random Forest \dotfill\begin{minipage}[t]{1.0cm}
\textbf{0.988}{.}\end{minipage}.
.2 Other Random Forest Variants.
.3 One-Sided Selection \dotfill\begin{minipage}[t]{1.0cm}
0.805{.}\end{minipage}.
.3 Neighborhood Cleaning Rule \dotfill\begin{minipage}[t]{1.0cm}
\textbf{0.998}{.}\end{minipage}.
.3 Random over sampling \dotfill\begin{minipage}[t]{1.0cm}
0.801{.}\end{minipage}.
.3 SMOTE \dotfill\begin{minipage}[t]{1.0cm}
0.796{.}\end{minipage}.
.4 borderline 1 \dotfill\begin{minipage}[t]{1.0cm}
0.814{.}\end{minipage}.
.4 borderline 2 \dotfill\begin{minipage}[t]{1.0cm}
0.813{.}\end{minipage}.
.4 Tomek links \dotfill\begin{minipage}[t]{1.0cm}
0.799{.}\end{minipage}.
.4 ENN \dotfill\begin{minipage}[t]{1.0cm}
\textbf{0.998}{.}\end{minipage}.
.2 Other methods$^*$ \dotfill\begin{minipage}[t]{1.0cm}
$\sim$ 0.0{.}\end{minipage}.
}
\caption{A hierarchical depiction of different sampling methods and the models used for each. The number associated with each methods indicates the $F_1$ score. Other methods$^*$: Since the data is highly unbalanced, the other models mostly overfit and learn to always predict the most dominant class. These models include Naive Bayes, Naive Bayes One-Sided Selection, Naive Bayes Neighborhood Cleaning Rule, Naive Bayes Random over sampling, Naive Bayes SMOTE and its variants.}
\label{fig:schematic}
\end{figure}

\subsubsection{Random Over-sampling}
Both Random Over-sampling and Random Under-sampling are baseline methods to combat the skewness of the dataset and balance the class distribution. In case of over-sampling, we randomly duplicate the samples of the minority class to have roughly the same number of datapoints in both classes.

\subsubsection{Random Under-sampling}
Random Under-sampling is the case of randomly deleting data points from the dominant class until both classes have roughly the same size. This method might delete the datapoints in the decision boundary that are important in the process of decision making.

\subsubsection{SMOTE \cite{14}} is a re-sampling method that generates new “synthetic” datapoints of the minority class using interpolation between the current datapoints. This may cause adding new datapoints in the space of the majority class.

\subsubsection{Tomek links \cite{15}} is an under sampling method. Tomek links removes the borderline and noisy datapoints. It takes two samples, $E_i$ and $E_j$ and computes d($E_i$, $E_j$) as their distance. A ($E_i$, $E_j$) is a Tomek link if there is no sample $E_k$ that $d(E_i, E_k)<d(E_i,E_j$) or $d(E_j ,E_k)<d(E_i, E_j$). After finding the links, the datapoints from the dominant class are be removed.

\subsubsection{One-sided Selection \cite{16}} uses the combination of Tomek links and \textit{CNN (Condensed Nearest Neighbor Rule) }to find the safe samples and removes the unsafe samples from the majority class. Tomek links removes the samples near the border line and CNN removes the samples that are far from the border line. 

\subsubsection{Neighborhood Cleaning Rule \cite{17}} removes some datapoints from the majority class. It finds the three nearest neighbors of each datapoint; if its neighbors belong to the majority class and the datapoint is from the minority class, then the neighbors will be removed. It the neighbors belong to the minority class and the datapoint is from the majority class,  

\subsubsection{NearMiss \cite{18}}
NearMiss 1, 2 and 3 algorithms are under-sampling methods. \textit{NearMiss 1} chooses the datapoints from the majority class whose average distance to three closest datapoints in the minority class is the smallest. \textit{NearMiss 2} chooses the majority class datapoints whose average to all datapoints in the minority class is the smallest. \textit{NearMiss 3} selects a given number of majority class datapoints for each datapoint in the minority class.

\subsubsection{SMOTE-Tomek links \cite{19}}
Since SMOTE over-sampling might lead to over-fitting and Tomek links under sampling might remove important datapoints, the ensemble of these two methods provide better results. In SMOTE-Tomek links, we first over-sample the minority class with SMOTE and then under-sample both the majority and minority classes producing a more balanced dataset.
\subsubsection{SMOTE-ENN \cite{20}} is also the ensemble of SMOTE and ENN. SMOTE is used as the over-sampler for minority class and then ENN provides data cleaning for both classes.

\subsection{Error Metrics}
In order to measure the performance of the models, error metrics are required. To show that the results are not biased and will roughly remain the same with new data, the dataset is randomly split into two sets; training and test. The training set is fed to a machine learning model. The trained model is then used to predict violations on the test set which was intact during the training and its true target values are known. This approach helps us select a model which will have good performance on unseen data.

To split the dataset into train and test sets, \textit{3-fold cross validation} is used. The dataset is randomly split into three partitions and the prediction model is trained three times. During each training, two-third of the dataset is used as the training set and fed to the model and one-third as the test set. The aggregated results of the three runs on the model will be reported as the final result.

In this paper, five different error metrics have been used to measure the performance of the models: \textit{Accuracy}, \textit{Receiver Operating Characteristic (ROC)  curves}, \textit{Precision}, \textit{Recall} and \textit{$F_\beta$ score}. Let us first define the \textit{Confusion Matrix} \cite{confu} which will help to define the above-mentioned metrics.

\begin{figure}[!ht]
  \centering
      \includegraphics[width=0.6\textwidth]{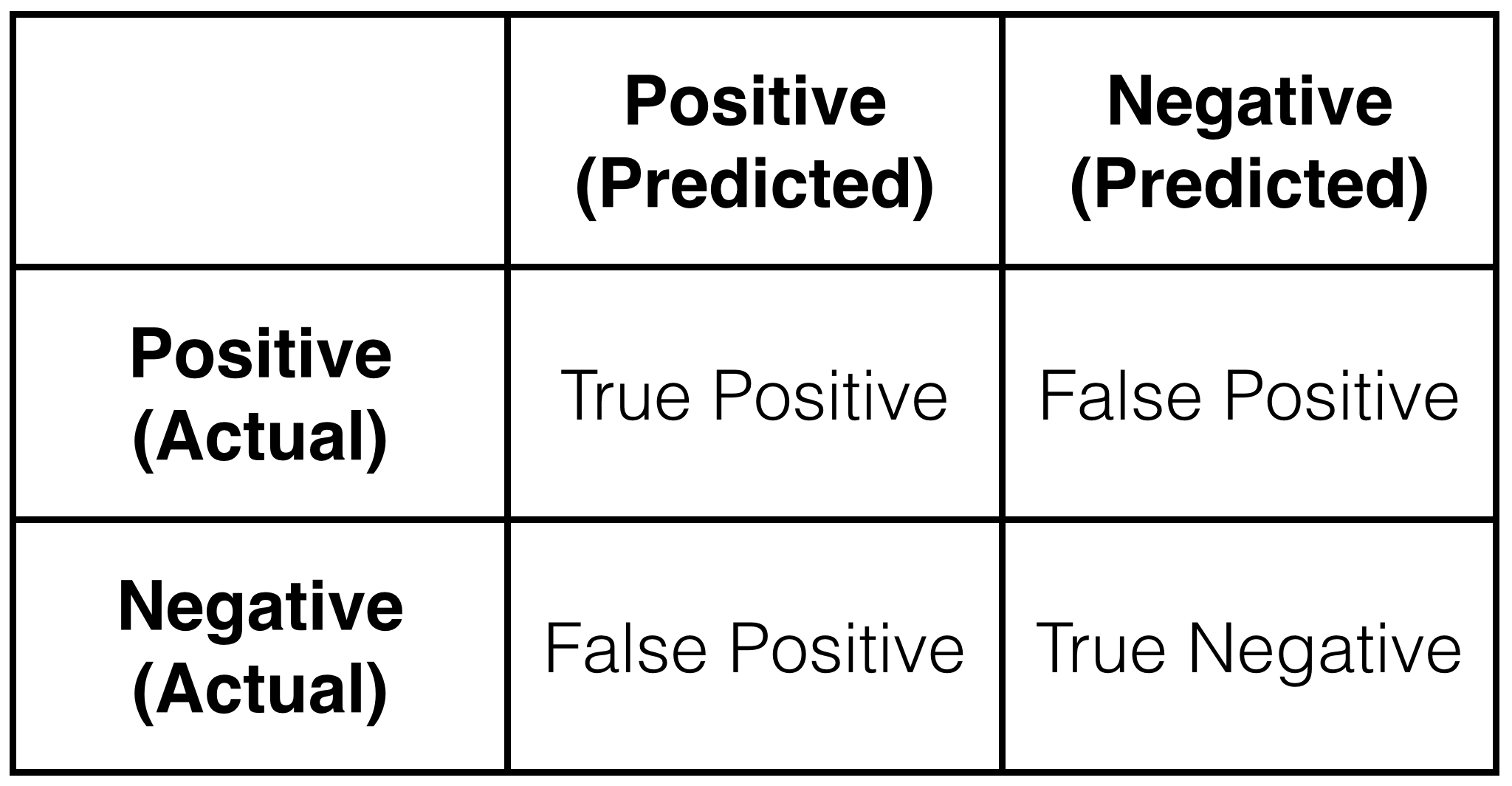}
  \caption{A confusion matrix: the table contains information about actual and predicted targets of a binary classifier.}
\label{fig:cm}
\end{figure}

As shown in Figure \ref{fig:cm}, a Confusion Matrix contains four values to describe the performance of a classification model: \textit{false positive}, \textit{false negative}, \textit{true positive}, and \textit{true negative}. False positive (resp. negative) is the number of mistakenly classified examples that are classified as 1 (resp. 0) where the actual targets are 0 (resp. 1). Similarly, true positive (resp. negative) is the number of correctly classified examples that are classified as 1 (resp. 0). Generally speaking, the first term (false or true) indicates if the classification result matches with the actual target. The second term (negative or positive) indicates the prediction of the classifier. Based on the confusion matrix, each of the metrics above are defined as follows:
\begin{align}
&accuracy = \frac{\mbox{True positives} + \mbox{True negatives}}{\mbox{\# of all examples}},\\
&precision = \frac{\mbox{True positives}}{\mbox{\# of all positive predictions}},\\
&recall = \frac{\mbox{True positives}}{\mbox{\# of all positive examples}},\\
&F_\beta = (1+\beta^2) \mbox{ . } \frac{\mbox{precision . recall}}{(\beta^2 \mbox{ . precision}) + \mbox{recall}}.
\end{align}

\begin{figure*}[t]
  \includegraphics[width=\textwidth]{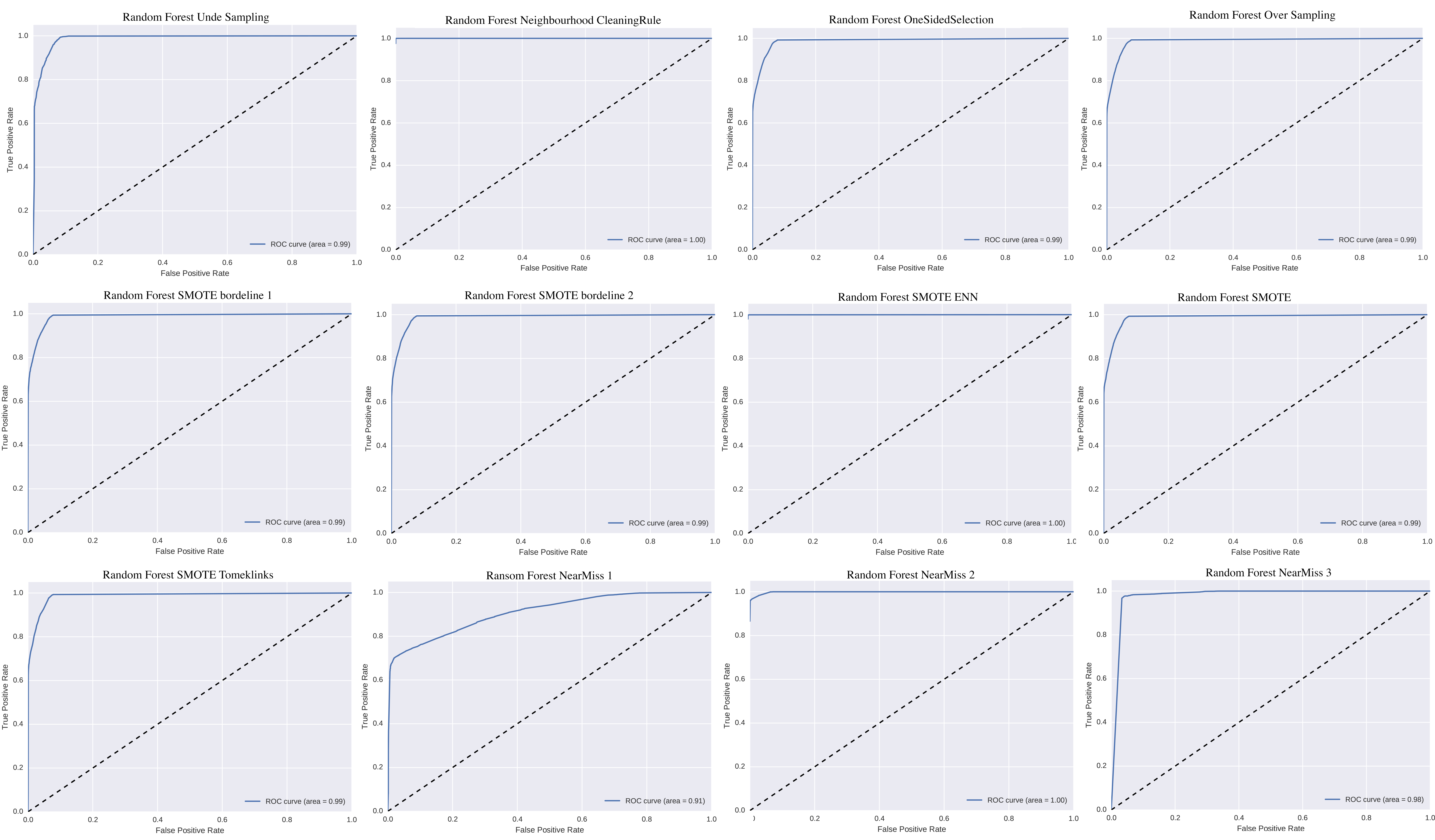}
  \caption{ROC curves of different sampling methods imposed on the random forest algorithm. ROC curves represent the performance of binary classifiers over different cut-off points of the algorithm. The area under the curve is considered as a single number presenting the trade-off between sensitivity (true positive rate) and specificity (true negative rate).}
\label{fig:rocs}
\end{figure*}

Finally, \textit{Receiver operating characteristic }(ROC) curve is a visualization of the performance of a binary classifier as its \textit{discrimination threshold} changes. The \textit{discrimination threshold} is the cut-off applied on the predicted probability of a test example that assigns it to a particular class. In an ROC curve, true positive rate and false positive rate are plotted on vertical and horizontal axis respectively.

\section{Results and Discussion}
\label{sec:results}
Table \ref{tbl:results} in Appendix A shows the results for different models and different sampling methods. A summary of the results showing $F_1$ score is presented in Figure \ref{fig:schematic}. The ROC curves of the Random Forest Classifier are also shown in Figure \ref{fig:rocs}. 

It is worth mentioning why error metrics other than accuracy are used. In skewed datasets, accuracy can not be a good error metric to find the best performing classifier. Two classes are available: 0.2\% of the samples are represented as violated class and 0.98\% of the samples are represented as unviolated. Consider a classifier that predicts there will be no violations. It has an accuracy of 0.98\% but Precision and Recall of zero. Thus, precision, recall and $f_\beta$ score will help us find the better performing algorithm.

\begin{figure}[!ht]
  \centering
      \includegraphics[width=0.7\textwidth]{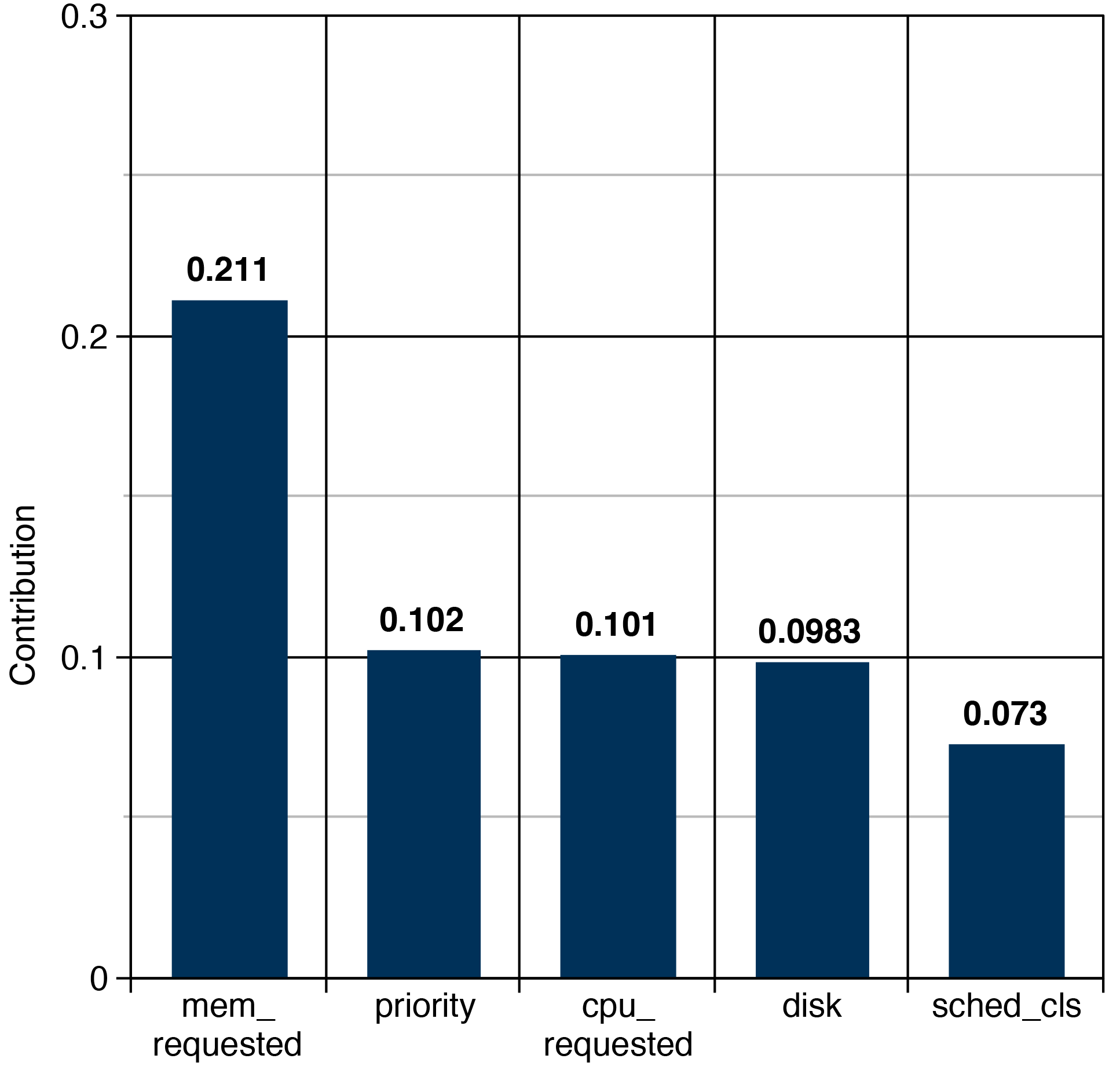}
  \caption{The average contribution of each feature based on the best trained random forest algorithm. The average is taken over all test examples.}
\label{fig:vis}
\end{figure}

The best performing model in terms of $F_1$ score is the random forest classification algorithm on the dataset re-sampled using the SMOTE + ENN method. According to the trained model, the five features shown in Figure \ref{fig:vis} are sorted based on their contributions in classification task. 

Our intuition is that the Random Forest has better performance because tree based classifiers  are less sensitive to class distributions. Thus, even with no re-sampling technique it has an acceptable performance (accuracy = 0.97\% and $f_1$ = 0.79). On the other hand, Naive Bayes classifiers are highly biased with class distribution and do not have any acceptable results without re-sampling techniques.

Among re-sampling methods, ensemble methods such as SMOTE-ENN, SMOTE-Tomek links and SMOTE-Borderline 1,2 had better results. SMOTE-ENN has better performance and our intuition is that because ENN removes more examples than the Tomek-Links, it provides more in depth data cleaning rule and removes any sample whose three nearest neighbors is miss-classified, which helps with better re-sampling the dataset. 

\section{Conclusion and Future Works}

The paper systematically compares the performance of two machine learning classification models on the task of SLA violation prediction. As discussed, in such a classification task, the data is skewed meaning that the number of violated tasks are much less than the number of unviolated ones. Consequently, the paper also explores several methods of handling unbalanced data. Results in section \ref{sec:results} suggest that, the random forest algorithm performs the best when SMOTE + ENN is used as the over-sampling method. Among other proposed models for tasks of SLA violation prediction or avoidance, our models are trained on a real world dataset which introduces new challenges that have been neglected in previous works, to the best of our knowledge. It is worth mentioning that the random forest model is not a black-box and the trained model is human-interpretable as the results suggests that \texttt{mem\_requested} is the most important feature in predicting violations.
Moreover, thanks to the relatively high speed of random forest, it can be used real-time.

Despite the impressive results achieved by random forest, one drawback of random forest is that it is not trivial to update the knowledge representation of the model based on the new coming examples. One of the future works might be to explore other models that can be easily updated when receiving more training data. Another remaining question in the area of SLA violation avoidance is how to take advantage of the prediction of classifier in order to avoid violation.
\FloatBarrier

\section*{Appendix A}
The full table of the results is presented in the following table.

\begin{table*}[h]
  \scalebox{0.9}{
  \begin{tabular}{c|ccccccc}
    \hline
    Model / Method & Accuracy & ROC area & Precision & Recall & $F_{0.5}$ & $F_1$ & $F_2$\\
    \hline \hline
NB (Gaus)                               & 0.9199   & 0.67     & 0.0       & 0.0    & 0.0                 & 0.0               & 0.0 \\ \hline
NB (Bern)                              & 0.9199   & 0.67     & 0.0       & 0.0    & 0.0                 & 0.0               & 0.0   \\ \hline
RF                                        & 0.9708   & 0.99     & 0.9353    & 0.6885 & 0.8728              & 0.7932            & 0.7269  \\ \hline
NB (Gaus)-under sampled                 & 0.5514   & 0.68     & 0.5293    & 0.8611 & 0.5735              & 0.6556            & 0.7652 \\ \hline
NB (Bern)-under sampled                & 0.6420   & 0.72     & 0.6788    & 0.5280 & 0.6421              & 0.5940            & 0.5526  \\ \hline
RF-under sampled                          & 0.9486   & 0.99     & 0.9361    & 0.9800 & 0.9446              & 0.9576            & 0.9709 \\ \hline
NB (Gaus)-near miss 1                   & 0.7594   & 0.75     & 0.7892    & 0.7023 & 0.7702              & 0.7432            & 0.7181 \\ \hline
NB (Bern)-near miss 1                  & 0.7159   & 0.71     & 0.8056    & 0.5629 & 0.7416              & 0.6627            & 0.5990  \\ \hline
RF-near miss 1                            & 0.8350   & 0.91     & 0.9546    & 0.7146 & 0.8945              & 0.8174            & 0.7525 \\ \hline
NB (Gaus)-near miss 2                   & 0.7506   & 0.74     & 0.7882    & 0.6908 & 0.7666              & 0.7363            & 0.7083  \\ \hline
NB (Bern)-near miss 2                  & 0.7078   & 0.71     & 0.8076    & 0.5519 & 0.7391              & 0.6557            & 0.5892 \\ \hline
RF-near miss 2                            & 0.8314   & 0.91     & 0.9629    & 0.7064 & 0.8977              & 0.8150            & 0.7462 \\ \hline
NB (Gaus)-near miss 3                   & 0.86712  & 0.88     & 0.9436    & 0.9086 & 0.9364              & 0.9258            & 0.9154  \\ \hline
NB (Bern)-near miss 3                  & 0.9129   & 0.72     & 0.9129    & 1.0    & 0.9291              & 0.9544            & 0.9812 \\ \hline
RF-near miss 3                            & 0.9747   & 0.98     & 0.9927    & 0.9826 & 0.9907              & 0.9876            & 0.9846  \\ \hline
NB (Gaus)-One-Sided Selection           & 0.9205   & 0.68     & 0.0       & 0.0    & 0.0                 & 0.0               & 0.0    \\ \hline
NB (Bern)-One-Sided Selection          & 0.9205   & 0.71     & 0.0       & 0.0    & 0.0                 & 0.0               & 0.0    \\ \hline
RF- One-Sided Selection                   & 0.9720   & 0.99     & 0.9503    & 0.6980 & 0.8862              & 0.8048            & 0.7372  \\ \hline
NB (Gaus) -Neighboorhood Cleaning Rule  & 0.8165   & 0.66     & 0.0206    & 0.0041 & 0.0115              & 0.0069            & 0.0049  \\ \hline
NB (Bern) -Neighboorhood Cleaning Rule & 0.8464   & 0.72     & 0.0       & 0.0    & 0.0                 & 0.0               & 0.0    \\ \hline
RF-Neighboorhood Cleaning Rule            & 0.9975   & 1.0      & 0.9985    & 0.9969 & 0.9981              & 0.9977            & 0.9972    \\ \hline
NB (Gaus) - Random over sampling        & 0.9124   & 0.68     & 0.0       & 0.0    & 0.0                 & 0.0               & 0.0    \\ \hline
NB (Bern) - Random over sampling       & 0.9124   & 0.71     & 0.0       & 0.0    & 0.0                 & 0.0               & 0.0    \\ \hline
RF - Random over sampling                 & 0.9688   & 0.99     & 0.9258    & 0.7058 & 0.8715              & 0.8009            & 0.7410  \\ \hline
NB (Gaus)- SMOTE                        & 0.9129   & 0.68     & 0.0       & 0.0    & 0.0                 & 0.0               & 0.0    \\ \hline
NB (Bern) - SMOTE                      & 0.9129   & 0.71     & 0.0       & 0.0    & 0.0                 & 0.0               & 0.0    \\ \hline
RF - SMOTE                                & 0.9690   & 0.99     & 0.9404    & 0.6908 & 0.8770              & 0.7965            & 0.7295  \\ \hline
NB (Gaus)-SMOTE borderline 1             & 0.9129   & 0.66     & 0.0       & 0.0    & 0.0                 & 0.0               & 0.0    \\ \hline
NB (Bern)-SMOTE borderline 1            & 0.9129   & 0.69     & 0.0       & 0.0    & 0.0                 & 0.0               & 0.0    \\ \hline
RF-SMOTE borderline 1                      & 0.9704   & 0.99     & 0.9189    & 0.7313 & 0.8740              & 0.8144            & 0.7624 \\ \hline
NB (Gaus) - SMOTE borderline 2           & 0.9129   & 0.66     & 0.0       & 0.0    & 0.0                 & 0.0               & 0.0    \\ \hline
NB (Bern) - SMOTE borderline 2          & 0.9128   & 0.69     & 0.0       & 0.0    & 0.0                 & 0.0               & 0.0    \\ \hline
RF - SMOTE borderline 2                    & 0.9696   & 0.99     & 0.9158    & 0.7318 & 0.8720              & 0.8135            & 0.7625  \\ \hline
NB (Gaus) - SMOTE Tomek links           & 0.9136   & 0.67     & 0.0       & 0.0    & 0.0                 & 0.0               & 0.0    \\ \hline
NB (Bern) - SMOTE Tomek links          & 0.9132   & 0.71     & 0.0       & 0.0    & 0.0                 & 0.0               & 0.0    \\ \hline
RF - SMOTE Tomek links                    & 0.9683   & 0.99     & 0.9452    & 0.6920 & 0.8808              & 0.7990            & 0.7312 \\ \hline
NB (Gaus) - SMOTE ENN                   & 0.7457   & 0.82     & 0.1701    & 0.7325 & 0.2010              & 0.2761            & 0.4409  \\ \hline
NB (Bern) - SMOTE ENN                  & 0.9317   & 0.79     & 0.0       & 0.0    & 0.0                 & 0.0               & 0.0    \\ \hline
RF - SMOTE ENN                            & 0.9988   & 1.0      & 0.9987    & 0.9972 & 0.9984              & 0.9980            & 0.9975   \\ \hline
    \hline
  \end{tabular} 
  }
  \caption{Full results of Naive Bayes and Random Forest classification algorithms and the sampling methods. NB and RF stand for Naive Bayes and Random Forest, respectively.}
\label{tbl:results}
\end{table*}

\end{document}